\documentclass{emulateapj}
\usepackage{epsfig}
\usepackage{apjfonts}
\usepackage{epsfig}

\newcommand{\lsim}{\lower0.6ex\vbox{\hbox{$ \buildrel{\textstyle
        <}\over{\sim}\ $}}}
\newcommand{\gsim}{\lower0.6ex\vbox{\hbox{$ \buildrel{\textstyle
        >}\over{\sim}\ $}}}

\newcommand{\hkpc}{ h^{-1}{\rm kpc}}
\newcommand{\Msun}{M_{\odot}}
\newcommand{\hMsun}{h^{-1} \Msun}
\newcommand{\hMpc}{h^{-1} {\rm Mpc}}

\newcommand{\lcdm}{$\Lambda$CDM~}

\newcommand{\beq}{\begin{equation}}
\newcommand{\eeq}{\end{equation}}

\shorttitle{The Assembly of Galaxy Clusters}
\shortauthors{Berrier et~al.}

\begin{document}

\title{The Assembly of Galaxy Clusters}

\author{
Joel C. Berrier,
Kyle R. Stewart,
James S. Bullock,
Chris W. Purcell,
Elizabeth J. Barton
}

\affil{Center for Cosmology, Department of Physics and Astronomy,
The University of California at Irvine, Irvine, CA 92697, USA}

\author{ Risa H. Wechsler}

\affil{  Kavli  Institute  for  Particle Astrophysics  and  Cosmology,
Physics Department,  and Stanford Linear  Accelerator Center, Stanford
University,\\ Stanford, CA 94305, USA }

\begin{abstract}

We study the formation  of fifty-three galaxy cluster-size dark matter
halos  ($M =  10^{14.0-14.76} \,  M_\odot$)  formed within  a pair  of
cosmological  \lcdm  N-body   simulations,  and  track  the  accretion
histories of cluster  subhalos with masses large enough  to host $\sim
0.1 L_*$ galaxies.  By  associating subhalos with cluster galaxies, we
find   the   majority   of   galaxies  in   clusters   experience   no
``pre-processing'' in  the group environment prior  to their accretion
into the cluster.   On average, $\sim 70 \%$  of cluster galaxies fall
into the cluster  potential directly from the field,  with no luminous
companions in their host halos at the time of accretion; and less than
$\sim  12 \%$  are accreted  as members  of groups  with five  or more
galaxies.  Moreover,  we find that cluster  galaxies are significantly
less  likely  to  have  experienced   a  merger  in  the  recent  past
($\lesssim6$ Gyr) than  a field halo of the  same mass.  These results
suggest  that local,  cluster processes  like  ram-pressure stripping,
galaxy  harassment,  or  strangulation   play  the  dominant  role  in
explaining the  difference between cluster and field  populations at a
fixed  stellar mass;  and that  pre-evolution or  past merging  in the
group  environment  is of  secondary  importance  for setting  cluster
galaxy properties  for most clusters.   The accretion times  for $z=0$
cluster  members are quite  extended, with  $\sim 20  \%$ incorporated
into the  cluster halo more than $7$  Gyr ago and $\sim  20 \%$ within
the last  $2$ Gyr.  By comparing the  observed morphological fractions
in cluster and field populations, we estimate an approximate timescale
for  late-type   to  early-type  transformation   within  the  cluster
environment to be $\sim 6$ Gyr.
\end{abstract}



\section{Introduction} \label{sec:intro}

Galaxy clusters are over-abundant in red, early-type galaxies compared
to  the field  population \citep[][]{Oemler1974,  B&O78, Dressler1980,
Dressler1997,   Treu2003,   Balogh2004,   Poggianti2006,   Capak2007}.
Approximately  $60 \%$ of  bright galaxies  located within  the virial
radii of cluster halos are  bulge-dominated (E + S0) compared to $\sim
30  \%$ of  similar luminosity  galaxies located  in  very low-density
environments \citep[][]{Whitmore_Gilmore91,Postman2005}.  The fraction
of  weakly  star-forming, early-type  galaxies  grows  with the  local
galactic density,  but even poor  groups show differences  compared to
the general population \citep[e.g.,][]{Postman_Geller84, Dressler1997,
zabludoff_mulchaey98, Tran2001, Finn2008}.

One  suggestion is  that ``pre-processing''  in the  group environment
prior  to cluster  formation is  important in  setting  the early-type
fraction                          in                          clusters
\citep[e.g.,][]{zabludoff_mulchaey98,Zabludoff2002}.      A    related
possibility  is that  galaxies in  clusters typically  experience more
mergers than field galaxies (prior  to their accretion), and that this
merger history bias plays  a role in explaining population differences
\citep[][]{T&T}.  Finally,  the fact that the overall  mix of galaxies
in clusters  by type  is known to  evolve with  redshift \citep{B&O78,
Ellingson2001,Tran2005,      Gerke2006,      Capak2007,      Coil2008,
Loh2008,Finn2008} suggests that the  internal cluster processes play a
major role  in setting the  differences between the cluster  and field
populations.  Here  we examine  the formation of  clusters in  a \lcdm
cosmological simulation in order to gain insight into these questions.

Galaxies  in clusters  and groups are  subject to a  number of
processes that may suppress star formation or change the morphology of
a    galaxy.      One    such    effect,     ram-pressure    stripping
\citep[][]{Gunn&Gott1972, Quilis_etal2000} seems to have been observed
directly in  Virgo \citep{chung07}, and simulations  suggest that this
process   should  operate   on   short,  $\sim   1$  Gyr,   timescales
\citep[][]{Tonnesen2007}.  Other processes of relevance include galaxy-galaxy
``harassment'' within the cluster potential \citep[][]{Moore1996}
and  cold gas ``strangulation''  \citep[][]{larson1980, kauffmann1993}, which cuts
of the gas supply for ongoing star formation in cluster galaxies.

By  incorporating subsets  of  these expected  cluster processes  into
semi-analytic  plus N-body  models, several  groups  have investigated
cluster   galaxy   population  trends   in   the   context  of   \lcdm
\citep[][]{Balogh2000,            Benson2001,            Diaferio2001,
Okamoto&Nagashima2001, Springel2001}.   Broadly speaking, these models
have been successful in producing the general behavior that early type
galaxies  are more  common  in clusters,  but  full agreement  between
theory and observation has yet to be achieved.  This is likely because
some  of  the  relevant   processes  (e.g.,   ram  pressure  stripping,
harassment)  have been  neglected,  and perhaps  because  some of  the
effects  that were  included (e.g.,  morphological  transformation via
mergers)  were  modeled  by  rough approximations.   Our  approach  is
related to these past theoretical  efforts, but different in its goal.
Specifically,  we aim to  quantify the  basic assembly  statistics for
cluster member halos using N-body  simulations and to present these as
a  basis  for  evaluating  scenarios  and  interpreting  observations.
Surely, many  of the  of the statistical  results presented  here were
implicitly  included  in  past  theoretical models  that  used  N-body
simulations  as  a  basis,  but  our  aim  is  to  present  the  \lcdm
predictions  as purely as  possible, without  obscuring them  with any
particular set of model assumptions for the baryon physics.

In  what follows  we use  a pair  of N-body  simulations to  track the
assembly  history  of  cluster-size  dark matter  halos.   In  section
\S~\ref{sec:methods}  we discuss  the simulations  and the  method for
finding   halos   and   subhalos.    We  present   our   findings   in
\S~\ref{sec:results}.    We  reserve  \S~\ref{sec:discussion}   for  a
discussion    of    potential    implications.    We    conclude    in
\S~\ref{sec:conclusion}.

\section{Methods}
\label{sec:methods}

We study the  formation histories of fifty-three $M  > 10^{14} \hMsun$
cluster-size dark matter halos  extracted from two cosmological N-body
simulations with comoving cubic volumes  of $120 \hMpc$ and $80 \hMpc$
on  a  side.   Each  simulation  corresponds to  a  flat  $\Lambda$CDM
cosmology with $\Omega_{\rm M} = 1 - \Omega_\Lambda = 0.3$, $h = 0.7$,
and $\sigma_8 = 0.9$ and  were performed using the Adaptive Refinement
Tree   (ART)   code   of   \citet{Kravtsov1997}.   As   discussed   in
\cite{Allgood2006}   and  \cite{Wechsleretal2006},   the   $80  \hMpc$
simulation followed the evolution of  $512^3$ particles with a mass of
$3.18 \times 10^8 \rm{\hMsun}$ and achieved a maximum force resolution
of  $1.2  \hkpc$.   The  $120  \hMpc$  simulation  \citep{Allgood2006}
followed  the evolution  of $512^3$  particles  with a  mass of  $1.07
\times  10^9 \rm{\hMsun}$  with  a maximum  force  resolution of  $1.8
\hkpc$.

We identify  halos in  the simulation using  a variation of  the Bound
Density Maxima  Algorithm \citep[BDM][]{Klypinetal1999a}, specifically
adopting   the  methods   outlined  in   \cite{Kravtsovetal2004}.   As
described  in Stewart et  al. (2008)  \nocite{Stewartetal2007}, virial
radii and virial  masses for halos are set by  the radius within which
the average  density is $\Delta_{\rm  vir}$ times the mean  density of
the   universe  \citep[e.g.,][]{BryanandNorman1998}.   At   $z=0$  this
definition implies  a radius-mass relation of $R_{\rm  vir} \simeq 951
(M_{\rm vir}/10^{14} \hMsun)^{1/3}$.  {\em Subhalos} are defined to be
self-bound halos  with centers located  within the radius of  a larger
halo.  In  most cases, subhalo density profiles  become overwhelmed by
the larger halo's density field at a truncation radius, $R_t$, that is
smaller than  $R_{\rm vir}$.  The  truncation radius is defined  to be
the radius where  a halo's density profile flattens  to a value larger
than $-0.5$.  We define a halo's  mass, $M$, to be the minimum $M_{\rm
vir}$ and the mass within $R_t$.

We use  $48$ snapshot outputs  from the $80 \hMpc$  simulation, spaced
roughly equally in expansion factor  $a = 1/(1+z)$ back to $z=21.6$ to
generate merger histories for halos and subhalos.  For the $120 \hMpc$
simulation we  used $91$ snapshot  outputs, spaced roughly  equally in
expansion factor  back to a  $z=10.1$.  The merger trees  were derived
using     the     methods     discussed     in     \cite{Allgood2006},
\cite{Wechsleretal2006},     and     Stewart     et     al.     (2008)
\nocite{Stewartetal2007}.  When halos  are accreted into larger halos,
we record the mass at this time and label it $M_{in}$.  For halos that
are not subhalos (field halos) we set $M_{in} = M$.  We use this mass,
$M_{in}$, as  a proxy  for luminosity in  defining our  cluster galaxy
samples (see  below).  Once a halo  becomes a subhalo,  we continue to
track its  mass as it evolves  within the larger  halo.  When subhalos
fall below  a critical mass,  $M_{cr} < M_{in}$, we  explicitly remove
them from our catalogs, and assume  that any galaxy it was hosting has
fallen out of  the observational sample (either because  it has lost a
significant  fraction  of  luminous   mass  or  because  it  has  been
disrupted).  This  choice for  $M_{cr}$ allows us  to define  a sample
cleanly at  a mass  scale where  our halo finder  is complete  and our
simulations are not strongly affected by over-merging.

In what follows  we will assume that halos  and subhalos with $M_{in}$
larger than a specific threshold will host one galaxy at their center.
If  a  halo  contains  one  or  more  subhalos,  it  is  assigned  one
``central'' galaxy in addition to one galaxy for each of its subhalos.
The term {\em host} is used to describe the largest halo that contains
a galaxy.   A halo need not  contain a subhalo  in order for it  to be
classified as  a {\em host}, it  simply needs to be  massive enough to
contain a galaxy at its center.  A subhalo cannot be a host.  Finally,
we  term {\em  field} halos  to be  all halos  that are  not contained
within  a larger  halo.  Thus,  by  definition field  halos cannot  be
subhalos.

By associating galaxies with subhalos larger than a critical mass {\em
at the time of their accretion}, we are adopting a strategy similar to
that     used     successfully     by    \cite{Conroyetal2006}     and
\cite{Berrieretal2006}.  These authors were able to reproduce both the
large-scale  and  small-scale  clustering  statistics of  galaxies  by
assuming a  monotonic relationship between the luminosity  of a galaxy
and the maximum  circular velocity that its halo had  when it is first
accreted    into   a   larger    halo   \citep[see][for    a   similar
approach]{wang2006}.

Our  primary population  of  cluster galaxies  is  defined by  setting
$M_{in} >  10^{11.5}$ and $M_{cr}  > 10^{11.0} \,  \hMsun$.  Averaging
over both simulation volumes, this choice defines a sample with number
density $n_g =  0.012 \, h^{3} {\rm Mpc}^{-3}$.   Matching this number
density to the  the Sloan Digital Sky Survey  (SDSS) r-band luminosity
function \citep{Blanton2003},  we estimate that  these cuts correspond
to a  galaxy population  with an r-band  magnitude brighter  than $M_r
\simeq  -18.5$ for  $h=0.7$.   This is  comparable  to the  luminosity
ranges     used     in     most     cluster     morphology     studies
\citep[e.g.,][]{Dressler1980,  Dressler1997,  Postman2005}. Note  that
our results are insensitive to the precise choice of $M_{cr}$. We have
redone the analysis  described below and using an  $M_{cr}$ value that
differs by  a factor of $\sim  3$ from our fiducial  $M_{cr}$ and find
virtually identical results.

Twelve of our  cluster halos are taken from  the higher resolution $80
\, \hMpc$ simulation.  For  these high-resolution clusters, we explore
the accretion histories of a second set of lower-mass cluster subhalos
with $M_{in}  > 10^{11.0}$ and  $M_{cr} > 10^{10.3} \,  \hMsun$.  This
sample has a number density of  $n_g = 0.042 \, h^{3} {\rm Mpc}^{-3}$,
which, with $h=0.7$, corresponds  to minimum luminosity of $M_r \simeq
-16.4$ using  the SDSS luminosity  function \citep{Blanton2003}.  This
sample will  be used to test  the dependence of our  results on sample
selection  and  should  represent  galaxies  that  are  comparable  to
(although somewhat  fainter than) the observational  sample studied by
\cite{Treu2003}.

Of  course,  these  estimated  cluster galaxy  luminosities  are  only
approximate.  We certainly don't  expect that a simple mapping between
mass and luminosity will hold in detail, especially within the cluster
environment.  However,  as we  show below, our  results do  not depend
strongly on  the adopted mass  cut.  This suggests that  the uncertain
mapping  between  individual   halo  masses  their  associated  galaxy
luminosities should not hinder the interpretation of our results.

The  total sample  of $53$  cluster halos  have $z=0$  masses spanning
$M_{\rm clus} =  (1.0 - 5.8) \times 10^{14} \,  \hMsun$, with a median
mass of $1.48  \times 10^{14} \, \hMsun$.  The  total number of galaxy
subhalos in  this, our main sample  of $M_{in} >  10^{11.5} \, \hMsun$
subhalos is  $834$ ($\sim  15$ per cluster).   We also  explore trends
with cluster halo mass by  dividing our main cluster sample into three
mass bins  containing 18, 17, and  18 clusters each,  with mass ranges
that span $10^{14.0-14.13}$, $10^{14.13-14.24}$, and $10^{14.24-14.76}
\hMsun$,  respectively.   The  clusters  in these  respective  samples
contain  an average of  $\sim 9$,  $\sim 14$,  and $\sim  25$ galaxies
each.  The subset of $12$  high-resolution cluster halos we study have
$z=0$ masses that  span $M_{\rm clus} = (1.0 -  3.4) \times 10^{14} \,
\hMsun$.  These clusters host a  total of $643$ subhalos that meet our
$M_{in} > 10^{11.0} \hMsun$ criterion and have an average of $\sim 54$
galaxies per cluster.   The most massive cluster in  this sample hosts
$102$ galaxies.   A summary of the  clusters we study  along with some
properties of their accretion histories are given in Table 1.

In order to investigate the  assembly of more massive clusters we also
study clusters generated semi-analytically using the code described in
\citet[][see  also \citealt{ZB03}]{Zentner_etal05}.   This  model uses
the  extended Press-Schechter (EPS)  formalism, applying  the specific
implementation of \citet{SomervilleKolatt99} to produce mass accretion
histories for host halos of  a given present-day mass, and then tracks
the orbital evolution of each  tree's subhalos in their host potential
via analytic prescriptions governing dynamical friction and tidal mass
loss.   The  initial orbital  energy  and  impact  parameter for  each
satellite system  are drawn from probability  distributions yielded by
cosmological  N-body  simulations.   This algorithm  produces  several
statistics which match those  produced in well understood cosmological
simulations.  While not as accurate as our numerical simulations, this
computationally-inexpensive semi-analytic  model allows us  to explore
trends with cluster mass (from  $10^{13} \hMsun \leq M_{\rm clus} \leq
10^{15.3}  \hMsun$) with  thousands  of stochastically-generated  halo
formation histories.

\section{Results} 
\label{sec:results}

\subsection{The environmental history of cluster galaxies}
\label{sec:mass}

\begin{figure}
\epsscale{1.0}
\plotone{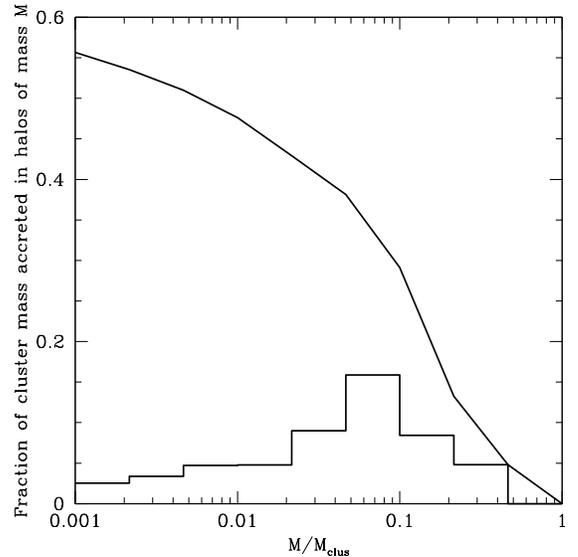}
\caption{Cumulative  and   differential  fraction  of   cluster  mass,
$M_{clus}$,  accreted  in halos  of  mass  $M/M_{\rm  clus}$ over  the
history of  the simulation.   This is an  average result based  on our
sample  of 53 clusters  with a  typical mass  of $M_{\rm  clus} \simeq
10^{14.2} \, \hMsun$.  }
 \label{fig:frac}
\end{figure}

Before presenting results on  the characteristics of cluster galaxies,
we first explore the {\em mass} assembly distribution for the clusters
themselves.  The  solid, monotonically increasing
line in  Figure \ref{fig:frac} shows  the mass
fraction of  our $z=0$ sample  of 53 clusters  that was built up from
accreting halos more massive than  $M$, where $M$ is the 
halo mass just prior to accretion into the cluster.
Note that we have normalized  the accreted halo
mass  $M$ relative  to  the  final cluster  mass  at $z=0$,  $M/M_{\rm
clus}$.  The histogram shows the corresponding differential distribution.
We see that the distribution peaks at $M/M_{\rm clus} \sim 0.1$.
This result is consistent with  the expectation that dark matter halos
of any size mass are built primarily from objects that  are $\sim 10 \%$
of the mass of the final halo \citep{Purcelletal07,Stewartetal2007,Zentner07}.
Note that more than half of the mass accreted from objects
large enough to host galaxies ($M/M_{\rm clus} \gtrsim 0.001 \, M_{\rm clus}$)
is accreted from group-scale systems with $M/M_{\rm clus} \gtrsim 0.1 \, M_{\rm clus}$.

%
%
%
%
%
%

\begin{figure}
\epsscale{1.0}
\plotone{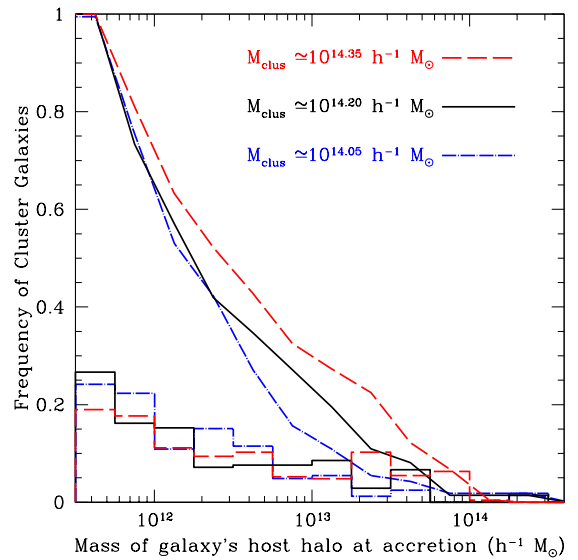}
\caption{  Cumulative and differential  fraction cluster  galaxies that
fell into their respective clusters as  part of a host halo of a given
mass.  Cluster galaxies  were  identified with  $M_{in} \ge  10^{11,5}
\hMsun$, corresponding to galaxies  brighter then $M_r \approx -18.5$.
The  red-dash,  black-solid,  and  blue-dot-dash lines  correspond  to
increasing  cluster  halo  mass  bins,  as labeled.   The  lines  that
increase monotonically towards lower accreted halo masses are the same
distributions presented cumulatively.  }
\label{fig:infallmass}
\end{figure}

Compare this  result to  Figure \ref{fig:infallmass}, which  shows the
fraction of $z=0$ cluster {\em galaxies} that were accreted as members
of host halos of a given mass.  The short-dashed (blue), solid (black)
and long-dashed (red)  lines correspond to clusters in  our three mass
bins,  centered  on $z=0$  masses  $M_{\rm  clus} \simeq  10^{14.05}$,
$10^{14.2}$, and  $10^{14.35} \hMsun$, respectively.  We  see that for
the typical cluster in our sample,  only $\sim 25 \%$ of the cluster's
galaxies were accreted as part of group-size objects with $M > 10^{13}
\, \hMsun$.   For the lowest and  highest mass subsamples  we see that
$\sim 15 \%$ and $\sim 30 \%$ of galaxies are accreted from group-mass
halos. The  results  are  not sensitive  to  the
selection of our $M_{cr}$ value.

The  previous two  figures show  that cluster  {\em mass}  assembly is
dominated  by the  most massive  (group-size) accretion  events, while
cluster {\em galaxy} assembly is dominated by lower-mass (galaxy-size)
halos.  This can  be understood by noting that  the number of galaxies
that a halo hosts does not  increase linearly with host mass.  A small
group ($M \sim 10^{12.5} \hMsun$) contains $\sim 10$ times the mass of
a single-galaxy halo ($M  \sim 10^{11.5} \hMsun$).  However, typically
a group halo of this mass  will host only $\sim 2-3$ galaxies that are
as bright  as the galaxies  associated with $10^{11.5}  \hMsun$ halos.
\citep[e.g.,][and  references  therein]{Berrieretal2006}.   This  means
that  a small  number of  group-size halos  can deposit  a significant
amount of  mass into a  cluster without contributing an  equally large
fraction of its galaxies.

%
%
%
%
%
%

\begin{figure}
\epsscale{1.0}
\plotone{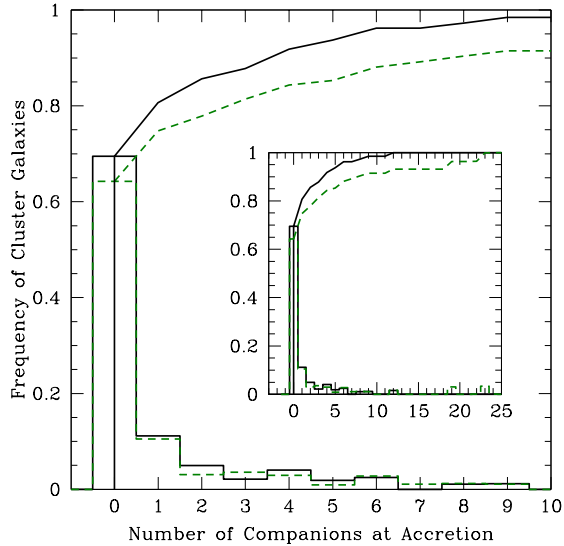}
\caption{  The  fraction of  surviving  cluster galaxies,  $N_g/N_{\rm
total}$, that  fell into their clusters  along with a  given number of
companions  in their  host dark  matter  halo.  The  black solid  line
corresponds to  our standard galaxy  sample ($M_r \lsim  -18.5$) while
the  green  dashed line  shows  the  same  distribution for  our  high
resolution sample  ($M_{in} \ge  10^{11} \hMsun$, $M_r  \lsim -16.4$).
The  inset box shows  the distribution  out to  the maximum  number of
companions found in our sample in  order to illustrate the tail of the
distribution.   Note that  ``companions'' do  not have  to  survive in
order to  be included  in the  count. We see  that, on  average, $\sim
70\%$ of cluster  galaxies fell into their clusters  directly from the
``field''  (with zero companions  in their  dark matter  halos), while
$\sim 80 \%$ fell in with 2 or fewer companions.}
\label{fig:comp}
\end{figure}

While the mass  of a galaxy's host at  accretion provides some insight
into the environment within which it formed and evolved before joining
the cluster, a more direct  measure of the environment can be obtained
by counting  the number of galaxies  that each host  halo contained at
the time of its  infall.  Figure \ref{fig:comp} shows the distribution
of the number  of {\em companions} each cluster  galaxy had within its
host halo at the time it was accreted into the cluster.  The companion
count includes any other galaxies  that existed within a galaxy's host
halo at  the time of  accretion.  We do  not require that  a companion
galaxy ``survives'' with $M > M_{cr}$  at $z=0$ in order to include it
in this  count --- it  simply must  have $M >  M_{cr}$ at the  time of
accretion into  the cluster.  A  companion count of zero  implies that
the galaxy was accreted as the only object in its halo ({\it i.e.,} it
was accreted from the field).  The two solid lines show the binned and
cumulative  distributions  for our  standard  sample  of 834  galaxies
within 53 clusters, and the pair of dark green dashed lines correspond
to  our sample  of 643  lower mass  galaxy halos  taken from  the $12$
high-resolution clusters.  For  our standard sample, approximately $70
\%$ of cluster galaxies were  accreted as the only galaxy within their
halo and $\sim 88 \%$ were accreted with fewer than $5$ galaxies.  The
numbers are similar for the sample that includes smaller galaxy halos.
In  this case,  $\sim  65 \%$  of  the cluster  galaxies are  accreted
directly  from  the field.  

%
%
%
%

%
%
%
\begin{figure}
\epsscale{1.0}
\plotone{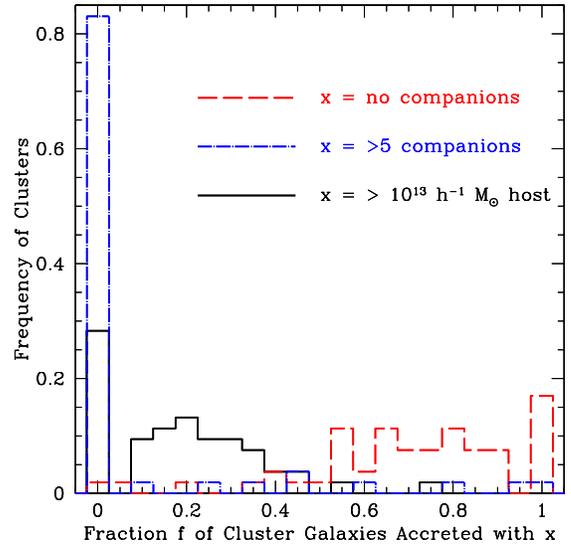}
\caption{  Distribution  of  galaxy  clusters with  a  given  assembly
history. The red-dashed line shows the frequency of simulated clusters
that have  a given  fraction $f$ of  their galaxies  accreted directly
from the  field (with no companions  in their host halos  as they fell
in, $f = N_0/N_{\rm total}$).  The red-dashed peak at $f=1$ shows that
$\sim 18 \%$ of our clusters have had 100\% of their galaxies accreted
directly from the field.  The blue-dot-dashed line shows the frequency
of clusters with a given fraction of their galaxies accreted with five
or more companions  in their host halos ($f  = N_{>5}/N_{\rm total}$).
The  large peak  at  $f=0$ shows  that  $\sim 83\%$  of our  simulated
clusters have  had none  of their galaxies  accreted from  groups with
five or more companions.  Finally,  the solid line shows the frequency
of simulated  clusters with  a given fraction  of their  galaxies that
were accreted as part of a  group-mass halo with $M > 10^{13} \hMsun$.
The broad bump around $f \sim 0.2$ shows that, typically, our clusters
have $\sim  10 - 40 \%$  of their galaxies accreted  from halos larger
than  the $\sim  10^{13} \hMsun$.   Here we  used our  standard galaxy
sample with $M_{in} \ge 10^{11.5} \hMsun$, $M_r \lsim -18.5$.}
\label{fig:hsplot}
\end{figure}
%
%
%

So far we  have considered only {\em averages}  for entire samples and
sub-samples of  our cluster-galaxy population.  We would  also like to
obtain  some indication of  the variation  in assembly  histories from
cluster  to cluster.  Figure  \ref{fig:hsplot} presents  statistics on
the  individual  ($M_{in}  >  10^{11.5}  \hMsun$)  galaxy  populations
subdivided into their respective  clusters.  The red-dashed line shows
the distribution of  our clusters that have a  given fraction ($f$) of
their  galaxies accreted  directly  from the  ``field'',  as the  only
object  in their  halos.  For  example, an  abscissa value  of $f=0.5$
implies  that 50 \%  of a  cluster's galaxies  were accreted  from the
field.  The peak in the red-dashed histogram at $f=1$ shows that $\sim
18  \%$ (10/53)  of our  clusters were  assembled entirely  from field
accretions.  The blue, short-dashed line shows the distribution of our
clusters that had  a fraction $f$ of their  galaxies accreted together
with $5$ or  more companions in their host halos at  the time of their
infall into the cluster.  The spike in the blue short-dashed histogram
at $f=0$ implies that $\sim 83 \%$ (44/53) of our clusters had none of
their  galaxies accreted  from a  group with  $5$ or  more companions.
Finally,  the  solid line  shows  the  frequency  of clusters  with  a
fraction $f$  of their galaxies accreted from  within group-mass halos
with $M \ge  10^{13} \hMsun$.  We see that  most clusters have between
$f =  10 \%$ and  $40 \%$ of  their galaxies accreted  from group-mass
halos.   However,  the  solid-line's   peak  at  $f=0$  shows  that  a
non-negligible fraction ($\sim 28 \%$, 15/53) of our clusters had none
of their galaxies accreted as members of group or cluster-size halos.

%
%

\subsection{Trends with Cluster Mass}

\begin{figure}
\epsscale{1.0}
\plotone{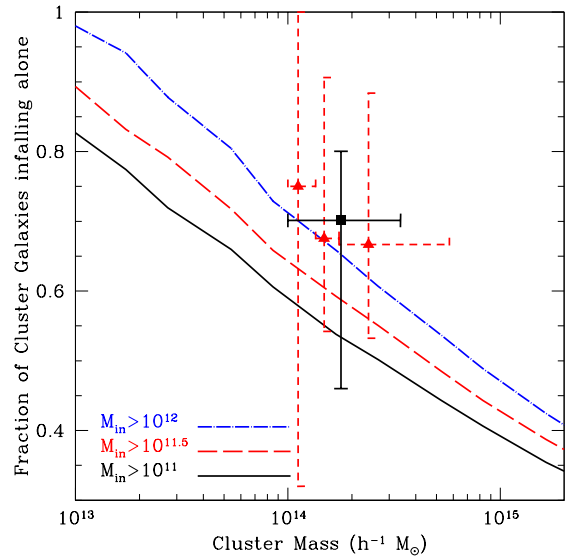}
\caption{ Fraction  of surviving cluster  galaxies that fell  into the
  cluster without a bright companion in their host halo at the time of
  accretion.  The three  triangle data points are taken  from our main
  sample of simulated,  where galaxy halos are defined  with $M_{in} >
  10^{11.5} \,  \hMsun$, $M_r  \le -18.5$.  The  square data  point is
  from  our high-resolution  sample,  $M_r \le  -16.4$, with  galaxies
  identified using our $M_{in}  > 10^{11} \, \hMsun$ criterion.  Error
  bars along  the y  axis for the  points reflect the  $68$ percentile
  range, and the error bars along the x-axis reflect the mass range of
  clusters  included  in  the  bin.   The  lines  correspond  to  data
  generated    by    the    semi-analytic   substructure    code    of
  \citet{Zentner_etal05}.  The  black (solid) line is for  a sample of
  galaxies with $M_{in} \ge 10^{11.0} \hMsun$ ($M_r \lsim -16.4$), the
  red  (dashed) line  is for  our standard  mass cuts  of  $M_{in} \ge
  10^{11.5}  \hMsun$ ($M_r  \lsim -18.5$),  and the  blue (dot-dashed)
  line is for  a sample with $M_{in} \ge  10^{12.0} \hMsun$.  Both the
  $M_{in}  >  10^{11.5}$  and  $M_{in}  > 10^{12.0}$  samples  have  a
  criteria of $M_{cr} \ge 10^{11.0} \hMsun$.  The $M_{in} > 10^{11.0}$
  sample matches our high resolution  sample with $M_{cr} \ge 2 \times
  10^{10.0}  \hMsun$.  Note  that  all  three samples  show  the  same
  trends. The only difference between them is that the trend is simply
  offset based on the mass of the sample.  }
\label{fig:alone}
\end{figure}

Figure \ref{fig:infallmass} shows that  more massive clusters are more
likely to  have had their  galaxies accreted from massive  host halos.
Figure  \ref{fig:alone} explores  the  issue of  mass dependence  more
fully.  Shown is the fraction  of surviving cluster galaxies that fell
into the cluster without a bright  companion in their host halo at the
time  of accretion  as  a function  of  the cluster  mass.  The  three
triangle data  points show  the median fractions  for three  mass bins
based on  our standard sample of  $53$ clusters with  galaxy halos set
via $M_{in}  > 10^{11.5} \, \hMsun$,  $M_r \lsim -18.5$  .  The square
data  point   is  from  our  high-resolution   sample,  with  galaxies
identified  using   our  $M_{in}  >  10^{11}   \,  \hMsun$  criterion,
corresponding to $M_r \lsim -16.4$.  Vertical error bars correspond to
the sixty-eighth  percentile range and the horizontal  error bars span
the mass range  of clusters included in the bins.   As expected, we do
see  a  slight  mass  trend  in  the  median,  with  the  fraction  of
field-accreted galaxies falling  from $\sim 75 \%$ to  $\sim 65 \%$ as
the cluster  mass increases from $\sim  10^{14} \, \hMsun$  to $\sim 3
\times 10^{14} \, \hMsun$, although the overall trend is weak compared
to the scatter from cluster-to-cluster at fixed mass.

In order  to explore how this  trend continues over  a broader cluster
mass  range,  we have  used  the \citet{Zentner_etal05}  semi-analytic
merger-tree  and substructure code  \citep[see also,][]{Purcelletal07}
to generate $1000$  cluster halo realizations at each  of $10$ cluster
halo  mass values  between  $M_{\rm  clus} =  10^{13}$  and $2  \times
10^{15} \hMsun$.   The three line types correspond  to three different
choices  for defining  the galaxy  samples, with  $M_{in}  > 10^{12}$,
$10^{11.5}$, and  $10^{11} \hMsun$, from  top-to bottom, respectively.
Galaxy subhalos  are counted  at $z=0$ as  surviving galaxies  if they
maintain  masses  larger  than  $M_{cr}  =  10^{11}$,  $10^{11}$,  and
$10^{10.5} \,  \hMsun$, respectively.   The long-dashed red  and solid
black lines  should correspond  to the red  triangle and  black square
N-body points,  respectively.  While the overall  normalization of the
semi-analytic estimates is low compared to the N-body result, the mass
trend  is  in  approximate   agreement,  with  more  massive  clusters
accreting  a smaller  fraction  of their  galaxies  directly from  the
field, with a mass scaling as $\sim M_{\rm clus}^{-0.2}$ over the mass
range plotted.  Note that we  expect even the most massive clusters to
have a  significant fraction of their galaxies  accreted directly from
the field.\\

%
%
%
%
%
%

%
%
%

\subsection{Cluster assembly with time}
\label{sec:time}

The  results  presented in  the  previous  section  suggest that  most
cluster galaxies experienced no evolution in a group environment prior
to their accretion into the  cluster, lending support to the idea that
internal cluster  processes are  responsible for the  differences seen
between cluster galaxies and those  in field environments.  If this is
so, then the distribution of time spent in the cluster environment can
provide  insight  into   the  timescales  required  for  morphological
transformation or the truncation of star formation.

\begin{figure}
\epsscale{1.0}
\plotone{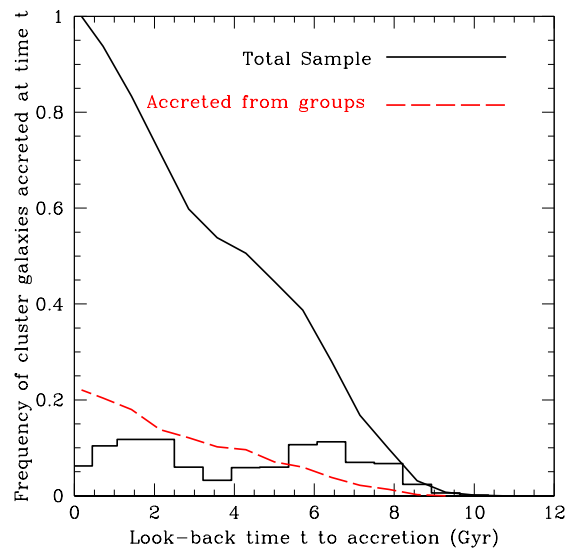}
\caption{ Differential and cumulative distributions of accretion times
for our primary sample, $M_r  \le -18.5$ of N-body cluster galaxies at
$z=0$.  The solid  black  line  shows the  distribution  of the  whole
sample. The red long-dashed  line shows the cumulative accretion times
for the subset of cluster  galaxies that were accreted from group-mass
halos  with $M>  10^{13} \hMsun$.   Note that  galaxies  accreted from
group  halos  are  biased  to  fall into  clusters  later  than  other
galaxies.  }
\label{fig:infalltime}
\end{figure}

Figure \ref{fig:infalltime} shows  the distribution of accretion times
for surviving  cluster galaxies  in our main  cluster sample.   We see
that the accretion rate has been fairly uniform over the past $\sim 8$
Gyr, with  the median  lookback time to  accretion at $\sim  4-5$ Gyr.
Interestingly,  as shown by  the red-dashed  line, galaxies  that were
acccreted as  part of group-mass halos  with $M >  10^{13} \hMsun$ are
biased  to   having  been  accreted   later  than  the   full  sample.
Specifically, the  median lookback time for galaxies  accreted as part
of a  group-mass system  is $\sim 3-4$  Gyr.  This is  not surprising,
since it takes  time for group-mass systems to  form in a hierarchical
universe.   Indeed,   we  suspect  that  this  fact   is  driving  the
double-peaked signature  in the accretion time histogram,  which has a
slight  dip at  $\sim 4$  Gyr.  We  do not  find a  trend  between the
accretion time distribution and  cluster mass within our sample. Table
1  lists  the  median   lookback  accretion  times  for  each  cluster
individually.

%
%
%
%
%
%

\begin{figure}
\epsscale{1.0}
\plotone{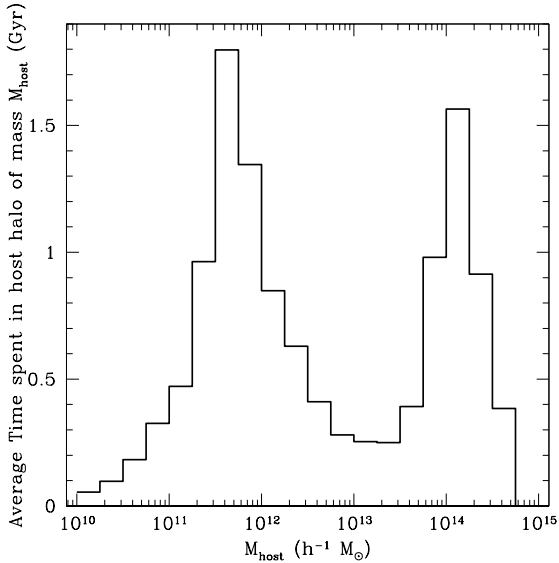}
\caption{  Average time that  $z=0$ cluster  galaxies (with  $M_{in} >
10^{11.5}  \hMsun$, $M_r  \lsim -18.5$)  spend in  host halos  of mass
$M_{\rm host}$.   Overall, cluster galaxies spend very  little time in
group-size of mass $10^{12.5-13.5} \hMsun$.  }
\label{fig:tir}
\end{figure}

Another way to gain insight into the history of cluster galaxies is to
quantify the  amount of time a galaxy  spends within a host  halo of a
given mass.  Figure  \ref{fig:tir} shows the mean time  that a cluster
galaxy has  spent in a  host halo of  a given mass, averaged  over the
history of the  simulation.  We see that this  time-weighted host mass
distribution  for cluster  galaxies is  {\em bimodal}  --  on average,
cluster galaxies have  spent time either in their  cluster or within a
galaxy-mass halo  prior to accretion. This  distribution is calculated
by examining the mass of each galaxy's dark matter halo at each of the
snapshots taken  from the  simulation.  The simplifying  assumption is
made that the halo spends all of its time at that given mass until the
next timestep.   The pronounced dip  in the middle shows  that cluster
galaxies tend  to {\em avoid} spending  time in groups.   Here we have
used the full  sample of cluster galaxies with  $M_{in} > 10^{11.5} \,
\hMsun$ to compute the average time that a cluster galaxy has spent in
a host halo of a given mass.  The peaks of the distribution are at the
single  galaxy-halo scale,  $\sim  10^{11.5} \,  \hMsun$,  and at  the
median cluster  scale $\sim 10^{14.2} \, \hMsun$.   Note that galaxies
can  spend some  time in  halos smaller  than our  $M_{in}$ threshold.
Specifically, these halos grew from a small mass to a mass larger than
$10^{11.5} \,  \hMsun$ before  being accreted.  It  is clear  that, on
average,  cluster galaxies spend  very little  time in  the group-mass
halos   between  $10^{12.5}$  and   $10^{13.5}  \,   \hMsun$. 

\begin{figure}
\epsscale{1.0}
\plotone{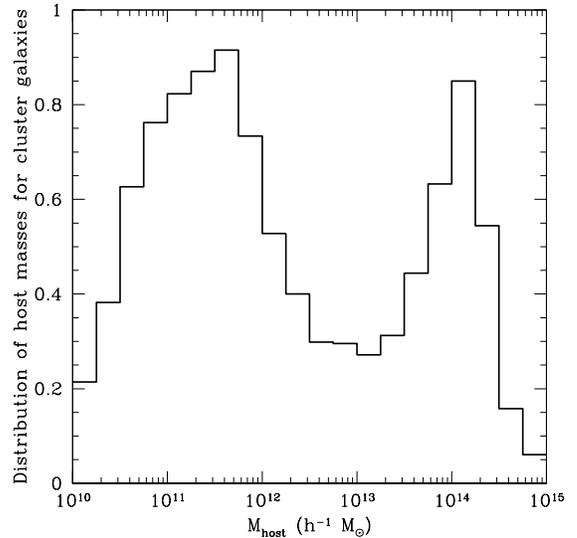}
\caption{ Fraction of $z=0$ cluster galaxies (with $M_{in} > 10^{11.5}
\hMsun$, $M_r \lsim  -18.5$) that have spent {\it any  time} in a host
of mass  $M_{\rm host}$. We allow  objects to appear  in multiple mass
bins as long as they have any time in a host halo of a given mass.}
\label{fig:mrh}
\end{figure}

Figure  \ref{fig:mrh} makes  a similar  point, but  in a  more extreme
fashion.  Here we  plot the fraction of galaxies  that have spent {\em
any} time  within a  halo within a  given mass  bin.  We use  the same
simplifying assumption  that the halo  remains at the same  mass until
the next time  step, therefore our ability to  determine whether halos
have  spent  anytime in  a  halo is  limited  by  our output  timestep
spacing.  In this  figure, a value of unity  implies that every galaxy
has spent at least some time in a halo of this size.

%
%
%
%
%
%

%
%
%
%
%
%

\begin{figure*}
\epsscale{1.0}
\plottwo{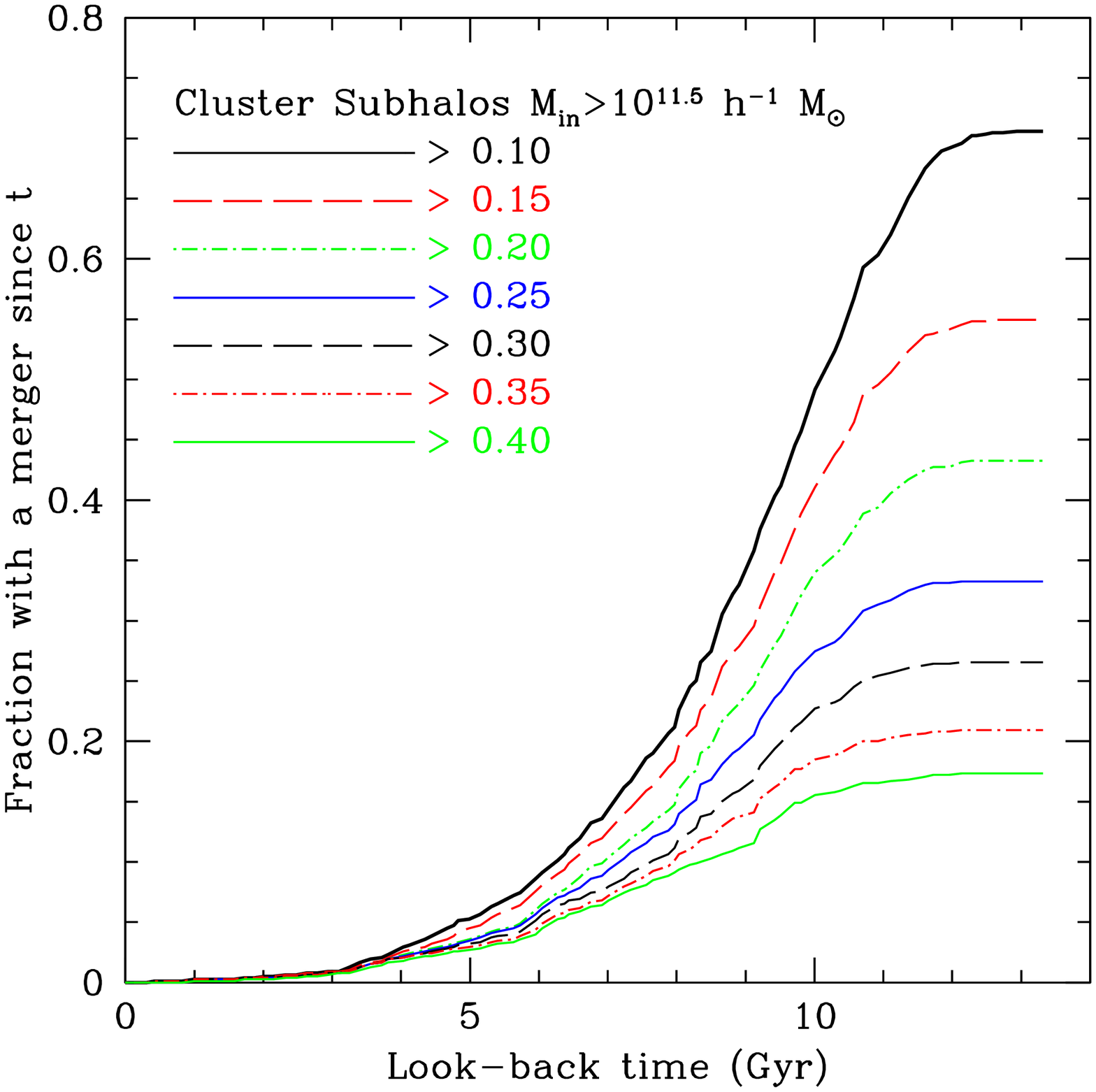}{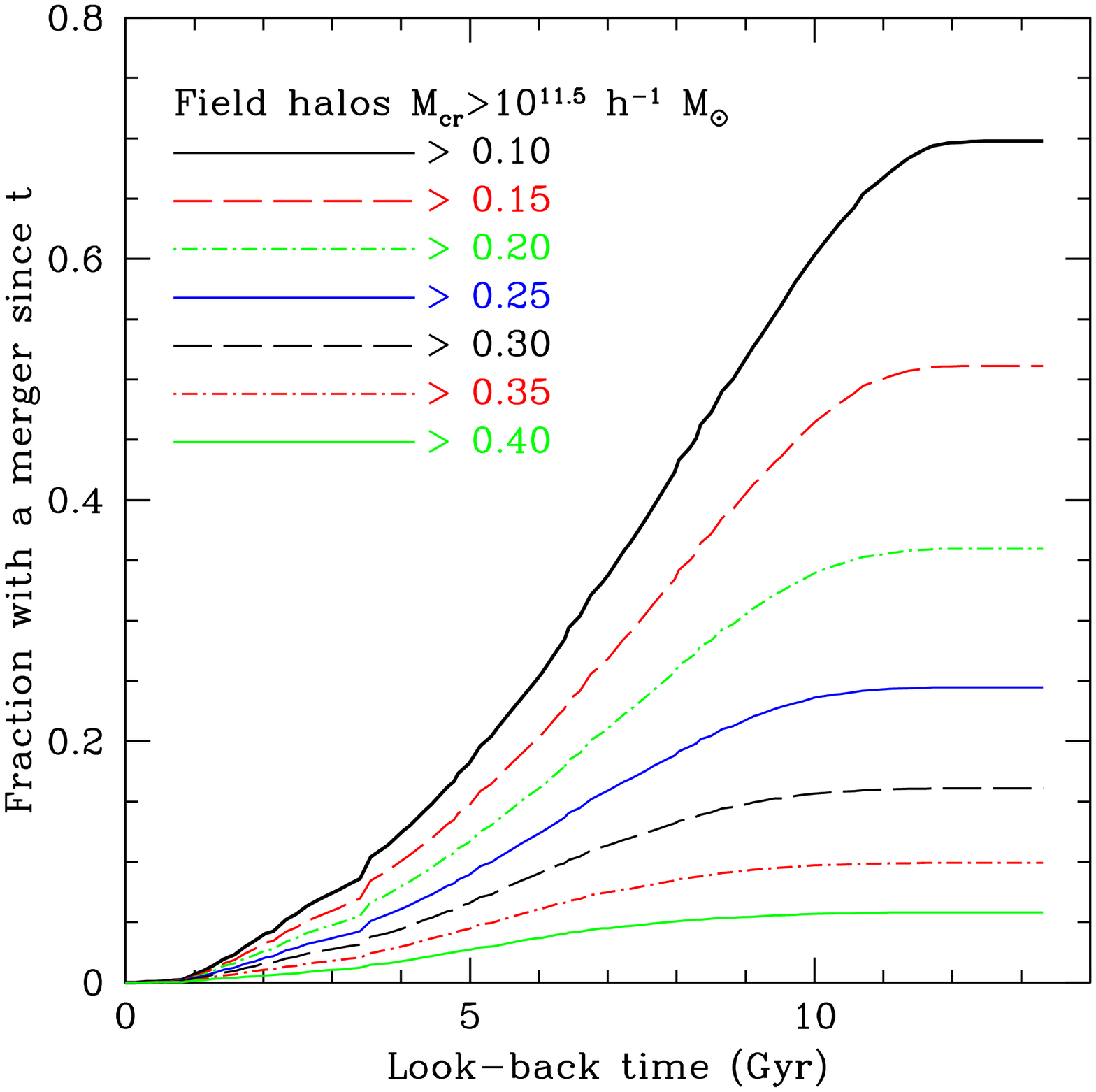}
\caption{Fraction of galaxy halos  that have undergone a merger larger
than a listed  fraction within a given look-back  time.  The {\em left
panel} is restricted to our  sample of cluster galaxies with $M_{in} >
10^{11.5}  \, \hMsun$  and the  {\em right  panel} includes  all field
halos  with $z=0$  masses larger  than  $ 10^{11.5}  \, \hMsun$.   The
labeled  line types  correspond to  different merger  ratio fractions,
$M/M_0$, where $M$ is the mass  of the merging object and $M_0$ is the
mass of the galaxy halo, with  $M_0 = M_{in}$ for cluster galaxies and
$M_0 = M$ at $z=0$ for field halos.  }
\label{fig:Mfrac}
\end{figure*}
%
%
%

\subsection{Cluster Galaxies and Merger Histories}

It is  common to  argue that spheroidal  galaxies are  associated with
halos that  have undergone significant mergers.  This  motivates us to
compare the merger histories of  field and cluster galaxies.  The left
panel  of  Figure \ref{fig:Mfrac}  shows  the  cumulative fraction  of
cluster  galaxies halos ($M_{in}  > 10^{11.5}  \, \hMsun$,  $M_r \lsim
-18.5$) that have have had a large merger since a given lookback time.
The right  panel shows the same  statistic, now for  {\em field} halos
with $M >  10^{11.5} \, \hMsun$.  The different  line types correspond
to  different mass-ratio  mergers: $m/M_0  = 1/10$  to $2/5$  from top
(solid) to bottom (dashed).  The mass ratio is defined to be the ratio
relative to $M_0 = M_{in}$ for cluster galaxies and relative to $M_0 =
M$ at $z=0$ for field halos.

At  large lookback  times  cluster galaxies  and  field galaxies  show
similar  results, at  least for  intermediate-size mergers.  The total
fraction of systems with $>1/10$  and $>1/5$ mergers in the last $\sim
10$ Gyr are  $\sim 75 \%$ and $\sim 40\%$,  respectively, for both the
cluster and  field populations.  The  cluster population has  a larger
fraction of systems that have  {\em ever} experienced {\em very} large
mergers.   For  example,  $\sim   25  \%$  of  cluster  galaxies  have
experienced something larger than a $0.3$ merger in the last $10$ Gyr,
compared to just $\sim 15 \%$ in the field.

Perhaps the most striking difference is that the cluster population is
much less  likely to have had  a {\em recent} ($<6$  Gyr) merger event
than galaxies in  the field.  The fraction of  cluster galaxies with a
significant merger in  the last $\sim 6$ Gyr is  less than $\sim 5\%$.
Given that half of cluster galaxies were accreted more than $\sim 4-5$
Gyr  ago (Figure  \ref{fig:infalltime}),  this general  result is  not
surprising.  We  expect the high-speed cluster  environment to greatly
reduce the likelihood for a  merger.  Large mergers are more likely to
occur in the field.

\section{Discussion}
\label{sec:discussion}

\subsection{Pre-processing}

Galaxy  groups provide an  important intermediate  environment between
the  field and high-density  clusters for  testing ideas  about galaxy
formation  and evolution  \citep{zabludoff_mulchaey98}. Indeed,  it is
possible  that pre-processing in  the group  environment prior  to the
assembly of galaxy  clusters is an important factor  in explaining why
cluster  galaxies  differ  significantly  from the  field  population.
Assuming  that $\Lambda$CDM  provides an  accurate description  of the
universe, the simulations presented  here allow us to characterize the
importance of groups in the global formation of galaxy clusters.

Note that with  our definition, groups and clusters  are defined using
the  standard ``virial''  over-density boundary  for host  dark matter
halos.   In this  case, most  galaxies in  the universe  {\em  do not}
reside in groups.  $\Lambda$CDM simulations suggest that only $\sim 10
\%$ of  $\sim $L$_*$  galaxy halos reside  within the virial  radii of
group     or     cluster     halos     \citep[e.g.,][and     references
therein]{Berrieretal2006}.   It  is interesting  then  to ask  whether
majority  of {\em  cluster}  galaxies were  assembled  from the  field
population or whether  they are biased to be  galaxies that evolved in
groups.

While Figure  \ref{fig:frac} demonstrates that  a significant fraction
of the {\em mass} in  clusters is accreted from group-size dark matter
halos, Figures  \ref{fig:infallmass} and \ref{fig:comp}  show that the
same is not true for {\em galaxies} in clusters.  Less than $17 \%$ of
cluster galaxies in $\sim 10^{14.2} \hMsun$ clusters fell in as a part
of a group with five or  more other galaxies.  Similarly, $\sim 25 \%$
were accreted as part of a halo more massive than $10^{13} \, \hMsun$.
This  finding suggests  that pre-processing  cannot play  the dominant
role in differentiating cluster galaxy populations from the field.  As
we discuss below, our  results suggest that global differences between
cluster and field populations  must be set by environmental influences
associated  with the clusters  themselves. 

It  is also  important  to emphasize  that  the pre-cluster  evolution
histories  of cluster  galaxies varies  significantly from  cluster to
cluster.  This  variation is demonstrated  in Figure \ref{fig:hsplot}.
For example,  while most of our  clusters ($44/53 \simeq  83 \%$) have
galaxy  populations that are  completely devoid  of objects  that were
accreted from  groups with five  or more companions, a  small fraction
($4/53 \simeq 7.5 \%$) of our  clusters have a majority ($f > 0.5$) of
their galaxies that fell into the cluster from groups of this kind.

\cite{zabludoff_mulchaey98}  have   suggested  that  cD   galaxies  in
clusters may form originally in the group environment.  Our results do
not prevent such a scenario from occurring in some cases, particularly
in those where  a large group-scale merger has  occurred recently.  As
seen in  Figures \ref{fig:tir} and  \ref{fig:mrh}, there are  always a
few galaxies  that evolve in groups. Additionally,  these figures only
include  galaxies  that survive  to  the  present.  Galaxies that  are
accreted into the  cluster and destroyed may provide  material for the
growth  of  cD  galaxies  and   is  consistant  with  the  results  of
\citet{Lin&Mohr2004}.

We also  find a  weak trend between  the fraction of  cluster galaxies
that  could  have  experienced  pre-processing  and the  mass  of  the
cluster.  For  the lowest  mass third of  our cluster  sample, $M_{\rm
clus} \simeq  10^{14.05} \hMsun$,  we see from  Figure \ref{fig:alone}
that $\sim 75  \%$ of cluster galaxies are  accreted directly from the
field.  This number drops to $\sim  65 \%$ in our largest mass sample,
$M_{\rm  clus} \simeq  10^{14.35} \hMsun$.   This trend  with  mass is
certainly real, however, the variation from cluster-to-cluster is much
stronger   than    the   mass   trend   itself.     The   results   of
\citet[][]{Weinmann2006}  suggest  that  the  early-type  fraction  of
cluster galaxies rises from $\sim 50 \%$ to $\sim 55\%$ in clusters of
mass  $\sim  10^{14}$ to  $10^{14.5}  \hMsun$.   It  is possible  that
pre-processing could play a role in driving this weak mass trend.

\subsection{Cluster Galaxy Merger Histories}

As we  showed in Figure  \ref{fig:Mfrac}, there are  clear statistical
differences between the merger histories of cluster galaxies and field
galaxy halos.  Nonetheless, it is unlikely that merger histories alone
can explain the morphology,  color, and spectral-type differences seen
between cluster galaxies and the  field.  For example, let us take $60
\%$  and  $30  \%$  as  characteristic  bulge-dominated  fractions  in
clusters  (within  $<  R_{\rm   vir}$)  and  the  field,  respectively
\citep[][]{Whitmore_Gilmore91,Postman2005}.    By   examining   Figure
\ref{fig:Mfrac} we see that  the field spheroid population potentially
could be associated directly with the $\sim 30 \%$ of field halos that
have experienced a $> 1/5$ merger  in the last $\sim 9$ Gyr.  However,
the $60\%$  cluster spheroid population  cannot be explained  with the
same set of assumptions: only $\sim 25 \%$ of cluster halos have had a
similar merger in the same period  of time, and only $\sim 45 \%$ have
{\em ever} experienced a $>1/5$ merger.

Given  that  merger histories  alone  cannot  explain the  differences
between cluster and field populations, we are forced to construct more
complicated scenarios  involving mergers to help  explain the observed
population  differences.  For  example,  galaxies  in  the  field  are
expected to  be surrounded by reservoirs  of baryons that  can cool to
reform a disk after a large merger.  Galaxies in clusters would likely
be stripped of this reservoir by the ambient cluster medium, making it
impossible for cluster  galaxies to accrete material to  reform a disk
after they  have fallen into the  cluster.  The right  panel of Figure
\ref{fig:Mfrac} shows  that we  could potentially associate  the $\sim
30\%$ of  bulge-dominated galaxies in  the field with the  $\sim 1/10$
mergers that  occurred within the last  $\sim 6$ Gyr  (and assume that
merger  remnants, older  than  $6$  Gyr, in  the  field have  reformed
disks).    However,  even  this   scenario  would   have  difficulties
explaining  the  cluster population:  only  $\sim  70  \%$ of  cluster
galaxies  have {\em ever}  experienced a  $>1/10$ merger  (left panel,
Figure \ref{fig:Mfrac}) and  only $\sim 35 \%$ were  accreted into the
cluster in the last $6$ Gyr.  Thus, obtaining more than a $\sim 25 \%$
spheroid  fraction  without   other  transformational  effects  (i.e.,
harassment) would seem difficult.

The  scenario discussed  above  is  similar to  that  adopted in  many
semi-analytic  models  \citep[e.g.,][]{Benson2001,Springel2001}, where
galaxies experience ``strangulation'' as their fresh gas supply is cut
off when they  fall into the cluster environment.   Not only does this
effect prevent the possible reformation of disks, but it can cause the
galaxy    to     redden    over    $\sim     1-3$    Gyr    timescales
\citep[][]{Poggianti1999,Balogh2000,  Ellingson2001}   and  thus  help
explain  the   morphological  {\em   and}  spectral  mix   of  cluster
populations.   It  should  be  noted, however,  that  observations  of
moderate redshift clusters suggest  that either two different physical
processes  or at least  two different  timescales are  responsible for
spectral    and     morphological    transformations    in    clusters
\citep{Poggianti1999}.

\subsection{Accretion Times and Galaxy Transformation}

The  timescales available  for  cluster galaxy  transformation can  be
garnered  directly  from   Figure  \ref{fig:infalltime}  (also  Figure
\ref{fig:tir}).  We expect  gravitational processes like harassment to
operate  over  timescales  similar  to  the  cluster  dynamical  time,
$1/\sqrt{G  \rho_{\rm  vir}}  \simeq   4$  Gyr.   Given  this,  Figure
\ref{fig:infalltime} shows that $\sim  70 \%$ of cluster galaxies have
been in  the cluster potential  long enough to  experience significant
dynamical perturbations.  In  order to transform a $\sim  70 \%$ field
disk population  to a  $\sim 40 \%$  cluster disk population,  we need
$\sim 3/7 \simeq 40 \%$  of accreted galaxies to be affected (assuming
spheroids remain  spheroids).  Figure \ref{fig:infalltime}  shows that
$\sim 40 \%$  of cluster galaxies have been  within the cluster longer
than  $\sim  6$  Gyr  ($\sim  1.5$ dynamical  times).   This  estimate
suggests that $\sim  6$ Gyr is a typical  transformation timescale for
cluster galaxies.

The timescale for strangulation is much shorter than a dynamical time,
and  should begin  to operate  as soon  as the  galaxy  encounters the
intra-cluster medium ($\lesssim 1$ Gyr).  We expect ram pressure to be
important  at higher  hot gas  densities than  strangulation  (i.e., at
smaller cluster-centric radii)  and therefore the associated timescale
for  ram pressure  to  act should  be  somewhat longer  than that  for
strangulation, perhaps  of order a  dynamical time.  Given  the infall
time distribution in Figure  \ref{fig:infalltime}, we expect that most
cluster galaxies will have been  affected to some degree by the cutoff
of fresh gas infall (e.g.,  $\sim  90\%$ with infall times $ > 1$ Gyr).
This  process  can alter  spectral  properties  without affecting  the
morphological mix.   A smaller fraction  have been in the  cluster for
$\sim 1$ dynamical time ($\sim 50  \%$ with infall times $> 4$ Gyr) --
long  enough  to   experience  significant  dynamical  (morphological)
transformations.

The  above discussions  provide some  qualitative evaluation  of ideas
that may explain  why the cluster galaxy population  is different from
the field  population.  A similar  analysis, involving a  more precise
characterization of radial dependencies  and merger histories, and the
evolution  of cluster  galaxy infall  times with  the redshift  of the
cluster, will be an important avenue for future investigation.\\

\section{Conclusions}
\label{sec:conclusion}

We use two cosmological  N-body simulations set within the concordance
\lcdm cosmology  to study the  formation of cluster-sized  dark matter
halos with masses spanning $M_{\rm clus} = 10^{14.0 - 14.76} \, \hMsun
$.  Our primary results are  based on tracking the merger histories of
galaxy subhalos  within these clusters.  These,  our cluster galaxies,
are picked to  have masses $M_{in} > 10^{11.5} \,  \hMsun$ at the time
they  first  fell  into   their  host  halos,  and  should  correspond
approximately to  $\sim 0.1$ L$_*$  galaxies.  Our conclusions  may be
summarized as follows.

\begin{itemize}

\item We find that the majority of cluster galaxies ($\sim 70 \%$) are
accreted directly  from the  field, as the  only objects  within their
dark matter halos at the time  of their infall into the cluster virial
radius.   A minority  ($\sim 25  \%$) were  accreted as  part of  $M >
10^{13} \hMsun$ group-mass halos, and  a small fraction ($\sim 12 \%$)
of  cluster galaxies  fell  into  their clusters  with  more than  $5$
galaxies in their halos at the time of accretion.

\item Cluster galaxy infall  histories show significant variation from
cluster to cluster.  For example,  9/53 ($\sim 17 \%$) of the clusters
in  our sample  were assembled  {\em entirely}  from the  accretion of
field galaxies.   However, 4/53  ($\sim 7 \%$)  of our clusters  had a
{\em majority} of their galaxies ($f > 0.5$) accreted from groups with
five or more companions.  Therefore, while on average cluster galaxies
tend to  be accreted from the  field, there are some  clusters that do
not follow this trend.

\item More massive clusters have  a smaller fraction of their galaxies
(at fixed luminosity) accreted directly from the field.  Approximately
$75 \%$ of $\sim 0.1 L^*$  galaxies in $M \simeq 10^{14} \hMsun$ halos
are accreted from  the field, compared to $\sim 65  \%$ of galaxies in
$M  \simeq  10^{14.35} \hMsun$  clusters.   The  scatter in  formation
histories from  cluster-to-cluster at  fixed mass is  more significant
than this mass trend.

\item  The  median lookback  time  to  accretion  for galaxies  within
clusters is $\sim 4.5$ Gyr, and  $\sim 85 \%$ of galaxies are accreted
between $1$ and $9$ Gyr ago.  By assuming that cluster galaxies
are accreted with a morphological mix similar to the field, we estimate
an approximate cluster-environmental 
transformation timescale of $\sim 6$ Gyr.

\item  Galaxy subhalos in  clusters are  significantly less  likely to
have had a  recent ($\lesssim 6$ Gyr) merger  than similar mass galaxy
halos in the field.  The merger fraction within the past $\sim 12$ Gyr
is $\sim 5 \%$ higher for the cluster subhalo population.

\end{itemize}

Taken  together, these  results suggest  that the  observed population
differences between  galaxies in clusters  and those in the  field are
driven  primarily by  internal cluster  processes.  Given  \lcdm  as a
basis,  merging  in  the  group  environment, or  any  other  type  of
pre-processing in galaxy groups prior to cluster assembly, cannot be a
major factor in setting the nearly two-to-one difference in early-type
fraction between clusters and the field.

Approximately half of an average cluster's population is accreted more
than $4$  Gyr ago ($\sim 1$  dynamical time).  This  allows ample time
for  gravitational processes  to  drive morphological  transformations
within  the  cluster  environment.   Moreover, interactions  with  the
intra-cluster medium  that remove gas  and suppress star  formation in
cluster galaxies  likely begin to  operate on even  shorter timescales
($\sim 1$ Gyr).  Therefore, most cluster galaxies ($ \sim 90 \%$) will
be affected by the cluster environment, at least to some degree.

While  our results  suggest that  pre-processing is  not  the dominant
mechanism  in  setting  galaxy  cluster  processes, we  do  find  most
clusters  have a  non-negligible fraction  of their  galaxies accreted
from  a group  environment.   This  is especially  true  for the  more
massive clusters  in our sample ($\sim 10^{14.35}  \hMsun$), for which
we find that $\sim 28 \%$ of their galaxies were accreted as part of a
halo larger than a group-mass  scale $M > 10^{13} \hMsun$.  Therefore,
some amount of preprocessing should occur.

As mentioned  in the introduction, important constraints  on the types
of  processes  that  act  to   shape  the  cluster  and  group  galaxy
populations  come  from  studies  at  intermediate  to  high  redshift
\citep{B&O78,Tran2005,Gerke2006,Capak2007,Finn2008}.     A    valuable
direction  of future  work will  be  to combine  the predicted  N-body
statistics  for cluster  halo assembly  time with  these results  as a
means to  constrain specific  scenarios for galaxy  transformation and
star formation suppression.

\acknowledgments 

The simulations were  run on the Seaborg machine  at Lawrence Berkeley
National  Laboratory (Project  PI:  Joel Primack).   We thank  Anatoly
Klypin for  running the simulation and  making it available  to us. We
acknowledge  Andrew   Zentner  for   the  use  of   his  semi-analytic
substructure code and for useful  comments on the manuscript.  We thank
Alison  Coil,  Jeff Cooke,  Asantha  Cooray,  Alan Dressler,  Margaret
Geller, Manoj  Kaplinghat, Andrey  Kravtsov, Jeremy Tinker,  and Frank
van den Bosch for useful  conversations.  JCB and JSB are supported by
NSF grant AST-0507916;  JCB, JSB, and EJB are  supported by the Center
for  Cosmology at  UC  Irvine.   RHW received  support  from the  U.S.
Department of Energy under contract number DE-AC02-76SF00515.

\bibliography{paper.bbl}

\begin{thebibliography}{51}
\expandafter\ifx\csname natexlab\endcsname\relax\def\natexlab#1{#1}\fi

\bibitem[{{Allgood} {et~al.}(2006){Allgood}, {Flores}, {Primack}, {Kravtsov},
  {Wechsler}, {Faltenbacher}, \& {Bullock}}]{Allgood2006}
{Allgood}, B., {Flores}, R.~A., {Primack}, J.~R., {Kravtsov}, A.~V.,
  {Wechsler}, R.~H., {Faltenbacher}, A., \& {Bullock}, J.~S. 2006, \mnras, 367,
  1781

\bibitem[{{Balogh} {et~al.}(2004){Balogh}, {Baldry}, {Nichol}, {Miller},
  {Bower}, \& {Glazebrook}}]{Balogh2004}
{Balogh}, M.~L., {Baldry}, I.~K., {Nichol}, R., {Miller}, C., {Bower}, R., \&
  {Glazebrook}, K. 2004, \apjl, 615, L101

\bibitem[{{Balogh} {et~al.}(2000){Balogh}, {Navarro}, \& {Morris}}]{Balogh2000}
{Balogh}, M.~L., {Navarro}, J.~F., \& {Morris}, S.~L. 2000, \apj, 540, 113

\bibitem[{{Benson} {et~al.}(2001){Benson}, {Frenk}, {Baugh}, {Cole}, \&
  {Lacey}}]{Benson2001}
{Benson}, A.~J., {Frenk}, C.~S., {Baugh}, C.~M., {Cole}, S., \& {Lacey}, C.~G.
  2001, \mnras, 327, 1041

\bibitem[{{Berrier} {et~al.}(2006){Berrier}, {Bullock}, {Barton}, {Guenther},
  {Zentner}, \& {Wechsler}}]{Berrieretal2006}
{Berrier}, J.~C., {Bullock}, J.~S., {Barton}, E.~J., {Guenther}, H.~D.,
  {Zentner}, A.~R., \& {Wechsler}, R.~H. 2006, \apj, 652, 56

\bibitem[{{Blanton} {et~al.}(2003){Blanton}, {Hogg}, {Bahcall}, {Brinkmann},
  {Britton}, {Connolly}, {Csabai}, {Fukugita}, {Loveday}, {Meiksin}, {Munn},
  {Nichol}, {Okamura}, {Quinn}, {Schneider}, {Shimasaku}, {Strauss}, {Tegmark},
  {Vogeley}, \& {Weinberg}}]{Blanton2003}
{Blanton}, M.~R., {Hogg}, D.~W., {Bahcall}, N.~A., {Brinkmann}, J., {Britton},
  M., {Connolly}, A.~J., {Csabai}, I., {Fukugita}, M., {Loveday}, J.,
  {Meiksin}, A., {Munn}, J.~A., {Nichol}, R.~C., {Okamura}, S., {Quinn}, T.,
  {Schneider}, D.~P., {Shimasaku}, K., {Strauss}, M.~A., {Tegmark}, M.,
  {Vogeley}, M.~S., \& {Weinberg}, D.~H. 2003, \apj, 592, 819

\bibitem[{{Bryan} \& {Norman}(1998)}]{BryanandNorman1998}
{Bryan}, G.~L. \& {Norman}, M.~L. 1998, \apj, 495, 80

\bibitem[{{Butcher} \& {Oemler}(1978)}]{B&O78}
{Butcher}, H. \& {Oemler}, Jr., A. 1978, \apj, 219, 18

\bibitem[{{Capak} {et~al.}(2007){Capak}, {Abraham}, {Ellis}, {Mobasher},
  {Scoville}, {Sheth}, \& {Koekemoer}}]{Capak2007}
{Capak}, P., {Abraham}, R.~G., {Ellis}, R.~S., {Mobasher}, B., {Scoville}, N.,
  {Sheth}, K., \& {Koekemoer}, A. 2007, \apjs, 172, 284

\bibitem[{{Chung} {et~al.}(2007){Chung}, {van Gorkom}, {Kenney}, \&
  {Vollmer}}]{chung07}
{Chung}, A., {van Gorkom}, J.~H., {Kenney}, J.~D.~P., \& {Vollmer}, B. 2007,
  \apjl, 659, L115

\bibitem[{{Coil} {et~al.}(2008){Coil}, {Newman}, {Croton}, {Cooper}, {Davis},
  {Faber}, {Gerke}, {Koo}, {Padmanabhan}, {Wechsler}, \& {Weiner}}]{Coil2008}
{Coil}, A.~L., {Newman}, J.~A., {Croton}, D., {Cooper}, M.~C., {Davis}, M.,
  {Faber}, S.~M., {Gerke}, B.~F., {Koo}, D.~C., {Padmanabhan}, N., {Wechsler},
  R.~H., \& {Weiner}, B.~J. 2008, \apj, 672, 153

\bibitem[{{Conroy} {et~al.}(2006){Conroy}, {Wechsler}, \&
  {Kravtsov}}]{Conroyetal2006}
{Conroy}, C., {Wechsler}, R.~H., \& {Kravtsov}, A.~V. 2006, \apj, 647, 201

\bibitem[{{Diaferio} {et~al.}(2001){Diaferio}, {Kauffmann}, {Balogh}, {White},
  {Schade}, \& {Ellingson}}]{Diaferio2001}
{Diaferio}, A., {Kauffmann}, G., {Balogh}, M.~L., {White}, S.~D.~M., {Schade},
  D., \& {Ellingson}, E. 2001, \mnras, 323, 999

\bibitem[{{Dressler}(1980)}]{Dressler1980}
{Dressler}, A. 1980, \apj, 236, 351

\bibitem[{{Dressler} {et~al.}(1997){Dressler}, {Oemler}, {Couch}, {Smail},
  {Ellis}, {Barger}, {Butcher}, {Poggianti}, \& {Sharples}}]{Dressler1997}
{Dressler}, A., {Oemler}, A.~J., {Couch}, W.~J., {Smail}, I., {Ellis}, R.~S.,
  {Barger}, A., {Butcher}, H., {Poggianti}, B.~M., \& {Sharples}, R.~M. 1997,
  \apj, 490, 577

\bibitem[{{Ellingson} {et~al.}(2001){Ellingson}, {Lin}, {Yee}, \&
  {Carlberg}}]{Ellingson2001}
{Ellingson}, E., {Lin}, H., {Yee}, H.~K.~C., \& {Carlberg}, R.~G. 2001, \apj,
  547, 609

\bibitem[{{Finn} {et~al.}(2008){Finn}, {Balogh}, {Zaritsky}, {Miller}, \&
  {Nichol}}]{Finn2008}
{Finn}, R.~A., {Balogh}, M.~L., {Zaritsky}, D., {Miller}, C.~J., \& {Nichol},
  R.~C. 2008, ArXiv:802.2282 [asstro-ph]

\bibitem[{{Gerke} {et~al.}(2007){Gerke}, {Newman}, {Faber}, {Cooper}, {Croton},
  {Davis}, {Willmer}, {Yan}, {Coil}, {Guhathakurta}, {Koo}, \&
  {Weiner}}]{Gerke2006}
{Gerke}, B.~F., {Newman}, J.~A., {Faber}, S.~M., {Cooper}, M.~C., {Croton},
  D.~J., {Davis}, M., {Willmer}, C.~N.~A., {Yan}, R., {Coil}, A.~L.,
  {Guhathakurta}, P., {Koo}, D.~C., \& {Weiner}, B.~J. 2007, \mnras, 376, 1425

\bibitem[{{Gunn} \& {Gott}(1972)}]{Gunn&Gott1972}
{Gunn}, J.~E. \& {Gott}, J.~R.~I. 1972, \apj, 176, 1

\bibitem[{{Kauffmann} {et~al.}(1993){Kauffmann}, {White}, \&
  {Guiderdoni}}]{kauffmann1993}
{Kauffmann}, G., {White}, S.~D.~M., \& {Guiderdoni}, B. 1993, \mnras, 264, 201

\bibitem[{{Klypin} {et~al.}(1999){Klypin}, {Gottl{\"o}ber}, {Kravtsov}, \&
  {Khokhlov}}]{Klypinetal1999a}
{Klypin}, A., {Gottl{\"o}ber}, S., {Kravtsov}, A.~V., \& {Khokhlov}, A.~M.
  1999, \apj, 516, 530

\bibitem[{{Kravtsov} {et~al.}(2004){Kravtsov}, {Gnedin}, \&
  {Klypin}}]{Kravtsovetal2004}
{Kravtsov}, A.~V., {Gnedin}, O.~Y., \& {Klypin}, A.~A. 2004, \apj, 609, 482

\bibitem[{{Kravtsov} {et~al.}(1997){Kravtsov}, {Klypin}, \&
  {Khokhlov}}]{Kravtsov1997}
{Kravtsov}, A.~V., {Klypin}, A.~A., \& {Khokhlov}, A.~M. 1997, \apjs, 111, 73

\bibitem[{{Larson} {et~al.}(1980){Larson}, {Tinsley}, \&
  {Caldwell}}]{larson1980}
{Larson}, R.~B., {Tinsley}, B.~M., \& {Caldwell}, C.~N. 1980, \apj, 237, 692

\bibitem[{{Lin} \& {Mohr}(2004)}]{Lin&Mohr2004}
{Lin}, Y.-T. \& {Mohr}, J.~J. 2004, \apj, 617, 879

\bibitem[{{Loh} {et~al.}(2008){Loh}, {Ellingson}, {Yee}, {Gilbank}, {Gladders},
  \& {Barrientos}}]{Loh2008}
{Loh}, Y.~., {Ellingson}, E., {Yee}, H.~K.~C., {Gilbank}, D.~G., {Gladders},
  M.~D., \& {Barrientos}, L.~F. 2008, ArXiv:802.3726 [astro-ph]

\bibitem[{{Moore} {et~al.}(1996){Moore}, {Katz}, {Lake}, {Dressler}, \&
  {Oemler}}]{Moore1996}
{Moore}, B., {Katz}, N., {Lake}, G., {Dressler}, A., \& {Oemler}, A. 1996,
  \nat, 379, 613

\bibitem[{{Oemler}(1974)}]{Oemler1974}
{Oemler}, A.~J. 1974, \apj, 194, 1

\bibitem[{{Okamoto} \& {Nagashima}(2001)}]{Okamoto&Nagashima2001}
{Okamoto}, T. \& {Nagashima}, M. 2001, \apj, 547, 109

\bibitem[{{Poggianti} {et~al.}(1999){Poggianti}, {Smail}, {Dressler}, {Couch},
  {Barger}, {Butcher}, {Ellis}, \& {Oemler}}]{Poggianti1999}
{Poggianti}, B.~M., {Smail}, I., {Dressler}, A., {Couch}, W.~J., {Barger},
  A.~J., {Butcher}, H., {Ellis}, R.~S., \& {Oemler}, A.~J. 1999, \apj, 518, 576

\bibitem[{{Poggianti} {et~al.}(2006){Poggianti}, {von der Linden}, {De Lucia},
  {Desai}, {Simard}, {Halliday}, {Arag{\'o}n-Salamanca}, {Bower}, {Varela},
  {Best}, {Clowe}, {Dalcanton}, {Jablonka}, {Milvang-Jensen}, {Pello},
  {Rudnick}, {Saglia}, {White}, \& {Zaritsky}}]{Poggianti2006}
{Poggianti}, B.~M., {von der Linden}, A., {De Lucia}, G., {Desai}, V.,
  {Simard}, L., {Halliday}, C., {Arag{\'o}n-Salamanca}, A., {Bower}, R.,
  {Varela}, J., {Best}, P., {Clowe}, D.~I., {Dalcanton}, J., {Jablonka}, P.,
  {Milvang-Jensen}, B., {Pello}, R., {Rudnick}, G., {Saglia}, R., {White},
  S.~D.~M., \& {Zaritsky}, D. 2006, \apj, 642, 188

\bibitem[{{Postman} {et~al.}(2005){Postman}, {Franx}, {Cross}, {Holden},
  {Ford}, {Illingworth}, {Goto}, {Demarco}, {Rosati}, {Blakeslee}, {Tran},
  {Ben{\'{\i}}tez}, {Clampin}, {Hartig}, {Homeier}, {Ardila}, {Bartko},
  {Bouwens}, {Bradley}, {Broadhurst}, {Brown}, {Burrows}, {Cheng}, {Feldman},
  {Golimowski}, {Gronwall}, {Infante}, {Kimble}, {Krist}, {Lesser}, {Martel},
  {Mei}, {Menanteau}, {Meurer}, {Miley}, {Motta}, {Sirianni}, {Sparks}, {Tran},
  {Tsvetanov}, {White}, \& {Zheng}}]{Postman2005}
{Postman}, M., {Franx}, M., {Cross}, N.~J.~G., {Holden}, B., {Ford}, H.~C.,
  {Illingworth}, G.~D., {Goto}, T., {Demarco}, R., {Rosati}, P., {Blakeslee},
  J.~P., {Tran}, K.-V., {Ben{\'{\i}}tez}, N., {Clampin}, M., {Hartig}, G.~F.,
  {Homeier}, N., {Ardila}, D.~R., {Bartko}, F., {Bouwens}, R.~J., {Bradley},
  L.~D., {Broadhurst}, T.~J., {Brown}, R.~A., {Burrows}, C.~J., {Cheng}, E.~S.,
  {Feldman}, P.~D., {Golimowski}, D.~A., {Gronwall}, C., {Infante}, L.,
  {Kimble}, R.~A., {Krist}, J.~E., {Lesser}, M.~P., {Martel}, A.~R., {Mei}, S.,
  {Menanteau}, F., {Meurer}, G.~R., {Miley}, G.~K., {Motta}, V., {Sirianni},
  M., {Sparks}, W.~B., {Tran}, H.~D., {Tsvetanov}, Z.~I., {White}, R.~L., \&
  {Zheng}, W. 2005, \apj, 623, 721

\bibitem[{{Postman} \& {Geller}(1984)}]{Postman_Geller84}
{Postman}, M. \& {Geller}, M.~J. 1984, \apj, 281, 95

\bibitem[{{Purcell} {et~al.}(2007){Purcell}, {Bullock}, \&
  {Zentner}}]{Purcelletal07}
{Purcell}, C.~W., {Bullock}, J.~S., \& {Zentner}, A.~R. 2007, \apj, 666, 20

\bibitem[{{Quilis} {et~al.}(2000){Quilis}, {Moore}, \&
  {Bower}}]{Quilis_etal2000}
{Quilis}, V., {Moore}, B., \& {Bower}, R. 2000, Science, 288, 1617

\bibitem[{{Somerville} \& {Kolatt}(1999)}]{SomervilleKolatt99}
{Somerville}, R.~S. \& {Kolatt}, T.~S. 1999, \mnras, 305, 1

\bibitem[{{Springel} {et~al.}(2001){Springel}, {White}, {Tormen}, \&
  {Kauffmann}}]{Springel2001}
{Springel}, V., {White}, S.~D.~M., {Tormen}, G., \& {Kauffmann}, G. 2001,
  \mnras, 328, 726

\bibitem[{{Stewart} {et~al.}(2008){Stewart}, {Bullock}, {Wechsler}, {Maller},
  \& {Zentner}}]{Stewartetal2007}
{Stewart}, K.~R., {Bullock}, J.~S., {Wechsler}, R.~H., {Maller}, A.~H., \&
  {Zentner}, A.~R. 2008, ApJ, accepted, ArXiv:711.5027 [astro-ph]

\bibitem[{{Tonnesen}(2007)}]{Tonnesen2007}
{Tonnesen}, S. 2007, New Astronomy Review, 51, 80

\bibitem[{{Toomre} \& {Toomre}(1972)}]{T&T}
{Toomre}, A. \& {Toomre}, J. 1972, \apj, 178, 623

\bibitem[{{Tran} {et~al.}(2001){Tran}, {Simard}, {Zabludoff}, \&
  {Mulchaey}}]{Tran2001}
{Tran}, K.-V.~H., {Simard}, L., {Zabludoff}, A.~I., \& {Mulchaey}, J.~S. 2001,
  \apj, 549, 172

\bibitem[{{Tran} {et~al.}(2005){Tran}, {van Dokkum}, {Illingworth}, {Kelson},
  {Gonzalez}, \& {Franx}}]{Tran2005}
{Tran}, K.-V.~H., {van Dokkum}, P., {Illingworth}, G.~D., {Kelson}, D.,
  {Gonzalez}, A., \& {Franx}, M. 2005, \apj, 619, 134

\bibitem[{{Treu} {et~al.}(2003){Treu}, {Ellis}, {Kneib}, {Dressler}, {Smail},
  {Czoske}, {Oemler}, \& {Natarajan}}]{Treu2003}
{Treu}, T., {Ellis}, R.~S., {Kneib}, J.-P., {Dressler}, A., {Smail}, I.,
  {Czoske}, O., {Oemler}, A., \& {Natarajan}, P. 2003, \apj, 591, 53

\bibitem[{{Wang} {et~al.}(2006){Wang}, {Li}, {Kauffmann}, \& {De
  Lucia}}]{wang2006}
{Wang}, L., {Li}, C., {Kauffmann}, G., \& {De Lucia}, G. 2006, \mnras, 371, 537

\bibitem[{{Wechsler} {et~al.}(2006){Wechsler}, {Zentner}, {Bullock},
  {Kravtsov}, \& {Allgood}}]{Wechsleretal2006}
{Wechsler}, R.~H., {Zentner}, A.~R., {Bullock}, J.~S., {Kravtsov}, A.~V., \&
  {Allgood}, B. 2006, \apj, 652, 71

\bibitem[{{Weinmann} {et~al.}(2006){Weinmann}, {van den Bosch}, {Yang}, \&
  {Mo}}]{Weinmann2006}
{Weinmann}, S.~M., {van den Bosch}, F.~C., {Yang}, X., \& {Mo}, H.~J. 2006,
  \mnras, 366, 2

\bibitem[{{Whitmore} \& {Gilmore}(1991)}]{Whitmore_Gilmore91}
{Whitmore}, B.~C. \& {Gilmore}, D.~M. 1991, \apj, 367, 64

\bibitem[{{Zabludoff}(2002)}]{Zabludoff2002}
{Zabludoff}, A. 2002, in Astronomical Society of the Pacific Conference Series,
  Vol. 257, AMiBA 2001: High-Z Clusters, Missing Baryons, and CMB Polarization,
  ed. L.-W. {Chen}, C.-P. {Ma}, K.-W. {Ng}, \& U.-L. {Pen}, 123--+

\bibitem[{{Zabludoff} \& {Mulchaey}(1998)}]{zabludoff_mulchaey98}
{Zabludoff}, A.~I. \& {Mulchaey}, J.~S. 1998, \apj, 496, 39

\bibitem[{{Zentner}(2007)}]{Zentner07}
{Zentner}, A.~R. 2007, International Journal of Modern Physics D, 16, 763

\bibitem[{{Zentner} {et~al.}(2005){Zentner}, {Berlind}, {Bullock}, {Kravtsov},
  \& {Wechsler}}]{Zentner_etal05}
{Zentner}, A.~R., {Berlind}, A.~A., {Bullock}, J.~S., {Kravtsov}, A.~V., \&
  {Wechsler}, R.~H. 2005, \apj, 624, 505

\bibitem[{{Zentner} \& {Bullock}(2003)}]{ZB03}
{Zentner}, A.~R. \& {Bullock}, J.~S. 2003, ApJ, 598, 49

\end{thebibliography}

\endpage
\appendix
\begin{table}
\begin{center}
\caption{Cluster Halo Properties}
\begin{tabular}{|l|c|c|c|c|c|c|c|c|}
\hline 
Cluster id & $M_{\rm clus} $ & $R_{\rm vir} $ & $N_{\rm grp}/N_{\rm total}$ & $N_{0}/N_{total}$ & $N_{\ge 5}/N_{total}$ & $T_{1/2}$ Mass & $T_{1/2}$ Galaxies & $N_{\rm grp}/N_{\rm total}$ \\
          &  ($10^{14} \hMsun$) &  ($\hMpc$) & ($>10^{11.5}$) & ($>10^{11.5}$) & ($>10^{11.5}$) & Gyr & Gyr & ($> 10^{11.0}$)\\
\hline
   1.120 &  5.86 &  1.71 &  15/53 &   34/53 &    6/53 &      7.1  &      2.4  &        NA  \\
   2.120 &  4.41 &  1.56 &  10/31 &   23/31 &    0/31 &      7.4  &      5.0  &        NA  \\
   3.120 &  3.80 &  1.48 &   3/23 &   19/23 &    0/23 &      7.4  &      4.9  &        NA  \\
   1.80  &  3.44 &  1.44 &   6/28 &   17/28 &    0/28 &      6.2  &      5.2  &    26/102  \\
   4.120 &  2.95 &  1.36 &   9/23 &   18/23 &    0/23 &      4.9  &      2.2  &        NA  \\
   5.120 &  2.77 &  1.34 &   5/26 &   14/26 &    0/26 &      5.2  &      5.8  &        NA  \\
   6.120 &  2.60 &  1.31 &   7/33 &   21/33 &    0/33 &      5.2  &      5.5  &        NA  \\
   7.120 &  2.57 &  1.30 &   7/21 &   14/21 &    0/21 &      3.4  &      1.8  &        NA  \\
   2.80  &  2.44 &  1.28 &   0/19 &   19/19 &    0/19 &      7.8  &      6.5  &     6/89   \\
   3.80  &  2.11 &  1.22 &   3/19 &   17/19 &    0/19 &      6.2  &      3.0  &    11/51   \\
   8.120 &  2.06 &  1.21 &   4/23 &    9/23 &   10/23 &      8.4  &      6.1  &        NA  \\
   9.120 &  1.94 &  1.19 &  23/41 &   21/41 &    9/41 &      6.6  &      4.5  &        NA  \\
  10.120 &  1.92 &  1.18 &   8/27 &   16/27 &    0/27 &      4.7  &      3.0  &        NA  \\
  11.120 &  1.88 &  1.17 &   6/22 &   18/22 &    0/22 &      8.1  &      5.8  &        NA  \\
   4.80  &  1.87 &  1.17 &   6/19 &   13/19 &    5/19 &      5.8  &      4.9  &    25/71   \\
  12.120 &  1.85 &  1.17 &   2/19 &   17/19 &    0/19 &      6.0  &      5.5  &        NA  \\
  13.120 &  1.80 &  1.16 &   2/17 &   15/17 &    0/17 &      5.8  &      4.5  &        NA  \\
  14.120 &  1.74 &  1.14 &  12/16 &    9/16 &    7/16 &      0.7  &      1.3  &        NA  \\
  15.120 &  1.71 &  1.14 &   0/17 &   17/17 &    0/17 &      8.7  &      5.7  &        NA  \\
  16.120 &  1.60 &  1.11 &   6/16 &    9/16 &    0/16 &      7.4  &      2.1  &        NA  \\
  17.120 &  1.59 &  1.11 &   7/15 &    8/15 &    0/15 &      2.4  &      3.0  &        NA  \\
   5.80  &  1.55 &  1.10 &   0/18 &   14/18 &    0/18 &      7.4  &      3.0  &     3/43   \\
   6.80  &  1.51 &  1.09 &   3/15 &   10/15 &    5/15 &      6.2  &      3.0  &     9/47   \\
  18.120 &  1.51 &  1.09 &   5/13 &    9/13 &    0/13 &      2.2  &      2.2  &        NA  \\
  19.120 &  1.49 &  1.09 &   0/10 &   10/10 &    0/10 &      6.5  &      7.1  &        NA  \\
   7.80  &  1.48 &  1.08 &   6/23 &   16/23 &    0/23 &      6.2  &      3.9  &    14/67   \\
  20.120 &  1.48 &  1.08 &   1/11 &    6/11 &    0/11 &      5.8  &      4.2  &        NA  \\
  21.120 &  1.48 &  1.08 &    3/8 &     7/8 &    0/8  &      3.6  &      1.8  &        NA  \\
  22.120 &  1.48 &  1.08 &    1/6 &     4/6 &    0/6  &      7.6  &      1.9  &        NA  \\
  23.120 &  1.45 &  1.08 &   1/10 &    8/10 &    0/10 &      7.3  &      2.1  &        NA  \\
   8.80  &  1.44 &  1.07 &   3/10 &    5/10 &    0/10 &      6.8  &      5.2  &     9/42   \\
   9.80  &  1.42 &  1.07 &    1/9 &     8/9 &    0/9  &      5.5  &      3.9  &     9/37   \\
  24.120 &  1.41 &  1.07 &   0/11 &    9/11 &    0/11 &      7.6  &      4.9  &        NA  \\
  25.120 &  1.39 &  1.06 &   3/12 &    8/12 &    0/12 &      7.4  &      7.1  &        NA  \\
  26.120 &  1.39 &  1.06 &    1/7 &     5/7 &    0/7  &      6.8  &      6.0  &        NA  \\
  27.120 &  1.34 &  1.05 &   3/13 &    0/13 &   13/13 &      5.5  &      5.8  &        NA  \\
  28.120 &  1.28 &  1.03 &    1/5 &     5/5 &    0/5  &      6.9  &      3.9  &        NA  \\
  29.120 &  1.27 &  1.03 &   0/12 &   10/12 &    0/12 &      6.0  &      5.0  &        NA  \\
  30.120 &  1.23 &  1.02 &    0/9 &     9/9 &    0/9  &      7.1  &      6.8  &        NA  \\
  31.120 &  1.23 &  1.02 &   3/15 &    8/15 &    0/15 &      6.6  &      2.4  &        NA  \\
  32.120 &  1.22 &  1.02 &   0/10 &   10/10 &    0/10 &      7.1  &      4.1  &        NA  \\
  33.120 &  1.16 &  1.00 &    0/8 &     6/8 &    0/8  &      7.4  &      5.3  &        NA  \\
  10.80  &  1.14 &  0.99 &    2/8 &     6/8 &    0/8  &      7.8  &      6.5  &     3/30   \\
  11.80  &  1.12 &  0.99 &   0/12 &   12/12 &    0/12 &      7.4  &      6.2  &     0/30   \\
  34.120 &  1.12 &  0.99 &    1/5 &     4/5 &    0/5  &      6.1  &      5.8  &        NA  \\
  35.120 &  1.12 &  0.99 &    1/7 &     6/7 &    0/7  &      5.8  &      1.8  &        NA  \\
  36.120 &  1.12 &  0.99 &    0/8 &     6/8 &    0/8  &      5.8  &      2.2  &        NA  \\
  37.120 &  1.10 &  0.98 &   0/12 &    5/12 &    7/12 &      5.2  &      4.4  &        NA  \\
  38.120 &  1.09 &  0.98 &    3/9 &     4/9 &    0/9  &      7.3  &      3.6  &        NA  \\
  39.120 &  1.08 &  0.98 &    4/9 &     2/9 &    7/9  &      8.4  &      0.7  &        NA  \\
  12.80  &  1.01 &  0.95 &    0/9 &     9/9 &    0/9  &      8.1  &      3.9  &     0/34   \\
  40.120 &  1.01 &  0.95 &    0/6 &     2/6 &    0/6  &      7.1  &      6.1  &        NA  \\ 	 
  41.120 &  1.01 &  0.95 &    0/9 &     9/9 &    0/9  &      8.5  &      4.7  &        NA  \\

\hline			          
\end{tabular}
\label{table:a}
\end{center}
{\small Note --  The first column denotes the  our id numbers, ordered
by mass within their  respective simulation box.  The number following
the decimal  point corresponds  to the box  size in  comoving $\hMpc$.
The second column  lists the cluster virial mass  in units of $10^{14}
\hMsun$.  The  third column  lists the cluster  virial radius  in $\rm
\hMpc$.   The fourth  column, $N_{\rm  grp}/N_{\rm total}$,  lists the
fraction of ($M_{in} >  10^{11.5} \hMsun$) galaxies that were accreted
in a group-size halos with $M  > 10^{13} \hMsun$.  The fifth and sixth
columns list the  fraction of galaxies that were  accreted as the only
object  in their  halo ($N_0/N_{\rm  total}$)  and with  five or  more
companions     in     their     halo     ($N_{>5}/N_{\rm     total}$),
respectively. Columns  seven and  eight show the  approximate lookback
times to the host halos accretion of  half of its mass and half of its
surviving substructure respectively.  Finally,  the last column is the
same  as  the fourth  column,  except now  we  track  the fraction  of
galaxies accreted in group-mass halos using our high-resolution sample
of $M_{in} > 10^{11} \hMsun$ galaxies.  }

\end{table}

\end{document}